\documentclass[twocolumn,aps,prl,showpacs,nofootinbib,nobibnotes]{revtex4}%
\usepackage{graphicx}
\usepackage{amsmath}
\usepackage{amsfonts}
\usepackage{amssymb}%
\providecommand{\U}[1]{\protect\rule{.1in}{.1in}}

\def\be{\begin{equation}}
\def\ee{\end{equation}}
\begin{document}
\title{Dynamic dissipative cooling of a mechanical oscillator in strong-coupling optomechanics}
\author{Yong-Chun Liu$^{1,2}$}
\author{Yun-Feng Xiao$^{1}$}
\altaffiliation{Corresponding author: yfxiao@pku.edu.cn; www.phy.pku.edu.cn/$\sim$yfxiao/index.html}
\author{Xingsheng Luan$^{2}$}
\author{Chee Wei Wong$^{2}$}
\altaffiliation{Corresponding author: cww2014@columbia.edu}
\affiliation{$^{1}$State Key Laboratory for Mesoscopic Physics and School of Physics,
Peking University, Beijing 100871, P. R. China}
\affiliation{$^{2}$Optical Nanostructures Laboratory, Columbia University, New York, NY
10027, USA}
\date{\today}

\begin{abstract}
Cooling of mesoscopic mechanical resonators represents a primary concern in
cavity optomechanics. Here in the strong optomechanical coupling regime, we
propose to dynamically control the cavity dissipation, which is able to
significantly accelerate the cooling process while strongly suppressing the
heating noise. Furthermore, the dynamic control is capable of overcoming
quantum backaction and reducing the cooling limit by several orders of
magnitude. The dynamic dissipation control provides new insights for tailoring
the optomechanical interaction and offers the prospect of exploring
macroscopic quantum physics.

\end{abstract}

\maketitle


One of the ultimate goals in quantum physics is to overcome the thermal noise,
so that quantum effects can be observed experimentally. A prominent example is
cavity optomechanics \cite{RevSci08,RevPhy09}, which enables not only the
fundamentally test of quantum theory and the exploration of quantum-classical
boundary, but also important applications in quantum information processing
and precision metrology. For these applications, the first crucial step is to
prepare the mechanical resonator into the quantum regime
\cite{GSNat11,GSNat11-2,SCNat12}. So far, numerous experiments have focused on
backaction cooling
\cite{CooNat06,CooNat06-2,CooNatPhys08,CooNatPhys09-1,CooNatPhys09-2,CooNatPhys09-3,CooNat10}
in the weak optomechanical coupling regime, holding potential for
ground-state preparation of mechanical resonators in the resolved sideband
condition \cite{PRL07-1,PRL07-2,PRA08}, along with backaction evading quantum
non-demolition measurements \cite{schwabNPhys2009,
aspelmeyerPNAS2011, braginskyJETPLett1978, braginskyBook}. Further step lies
in the strong coupling, essential for coherent quantum optomechanical
manipulations \cite{SCNat09,SCNat12,ST12,SCPRL08,SCPRA09,SCNJP10} and
electromechanical interactions \cite{GSNat10,TaylorPRL11}. However, till date,
strongly-coupled optomechanical cooling has predicted only
limited improvement over weak coupling due to the saturation
effect of the steady-state cooling rate \cite{SCPRL08,SCNJP08,PhasePRA09}.
Although strong coupling allows state swapping \cite{SCNat12,ST12}, it cools
the mechanical resonator only at a single instant in the Rabi oscillation
cycle. Thus it is urgent to overcome these limitations for cooling and
manipulating mesoscopic mechanical systems in the quantum regime.

For this purpose, in this Letter we show the dynamic tailoring of the cooling
and heating processes by exploiting the cavity dissipation, overcoming the saturation of the steady-state cooling rate. This greatly
accelerates the cooling process and thereby strongly suppresses the thermal
noise. Moreover, heating induced by swapping and interaction quantum
backaction are largely suppressed by periodic modulation of the cavity
dissipation, which breaks the fundamental limitation of backaction cooling.

\begin{figure}[tb]
\centerline{\includegraphics[width=\columnwidth]{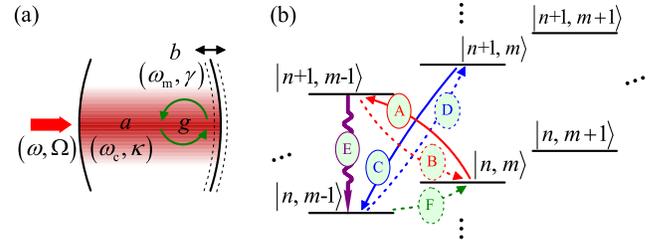}}
\caption{(color) (a) Sketch of a typical optomechanical system. (b) Level
diagram of the linearized Hamiltonian (\ref{HL}). $\left\vert n,m\right\rangle
$ denotes the state of $n$ photons and $m$ phonons in the displaced frame. The
solid (dashed) curves with arrows correspond to the cooling (heating)
processes.}%
\label{Fig1}%
\end{figure}

We consider a generic optomechanical system in which an optical cavity driven
by a laser is coupled to a mechanical resonance mode, as illustrated in Fig.
\ref{Fig1}(a). In the rotating frame at the driven laser frequency $\omega$,
the system Hamiltonian reads $H=-({\omega-\omega_{\mathrm{c}}}){a^{\dag}%
a}+{\omega_{\mathrm{m}}b^{\dag}b+ga^{\dag}a{(b+b^{\dag})+({\Omega{{a^{\dag}%
}+\Omega}}}}^{\ast}a{)}$ \cite{LawPRA95}, where ${a}$ (${b}$) represents the
annihilation operator for the optical (mechanical) mode with ${\omega
}_{\mathrm{c}}$ (${\omega_{\mathrm{m}}}$) being the corresponding angular
resonance frequency; $g$ denotes the single-photon optomechanical coupling
rate; ${\Omega}$ represents the driving strength. For strong driving, the
Hamiltonian can be linearized, with ${a\equiv a}_{1}+\alpha$, $b\equiv
b_{1}+\beta$. Here $a_{1}$ and $b_{1}$ describe the fluctuations around the
mean values $\alpha\equiv\langle a\rangle$ and $\beta\equiv\langle b\rangle$,
respectively. Neglecting the nonlinear terms, this yields the Hamiltonian
\begin{equation}
H_{L}=-\Delta^{\prime}{a_{1}^{\dag}a}_{1}+{\omega_{\mathrm{m}}{b_{1}^{\dag}%
b}_{1}+(Ga_{1}^{\dag}+G^{\ast}a_{1})(b_{1}+b_{1}^{\dag}}),\label{HL}%
\end{equation}
where ${\Delta}^{\prime}=\omega-{{\omega}_{\mathrm{c}}}+2|{G}|^{2}%
/{\omega_{\mathrm{m}}}$ is the optomechanical-coupling modified detuning, and
$G={\alpha g}$ describes the linear coupling strength. Taking the dissipations
into consideration, the system is governed by the quantum master equation
$\dot{\rho}=i[\rho,H_{L}]+\kappa\mathcal{D}[{a}_{1}]\rho+\gamma(n_{\mathrm{th}%
}+1)\mathcal{D}[{{b}_{1}}]\rho+\gamma n_{\mathrm{th}}\mathcal{D}[{{b_{1}%
^{\dag}}}]\rho$, where $\mathcal{D}[\hat{o}]\rho=\hat{o}\rho\hat{o}^{\dag
}-{({\hat{o}^{\dag}\hat{o}\rho+\rho\hat{o}^{\dag}\hat{o})/2}}$ denotes the
Liouvillian in Lindblad form for operator $\hat{o}$; $\kappa\equiv{\omega
}_{\mathrm{c}}/{Q}_{\mathrm{c}}$ ($\gamma\equiv{\omega}_{\mathrm{m}}%
/{Q}_{\mathrm{m}}$) represents the dissipation rate of the optical cavity
(mechanical) mode; $n_{\mathrm{th}}=1/(e^{\hbar{\omega_{\mathrm{m}}%
/k}_{\mathrm{B}}T}-1)$ corresponds to the thermal phonon number at the
environmental temperature $T$.

Figure \ref{Fig1}(b) displays the level diagram of $H_{L}$ and the coupling
routes among states $|n,m\rangle$ with $n$ ($m$) being the photon (phonon)
number in the displaced frame. We note that there are three kinds of heating
processes denoted by the dashed curves in Fig. \ref{Fig1}(b), corresponding to
swap heating ($B$), quantum backaction heating ($D$) and thermal heating
($F$). Suppressing thermal heating is the ultimate goal while swap heating and
quantum backaction heating are the accompanying effect when radiation pressure
is utilized to cool the mechanical motion. Swap heating emerges when the
system is in the strong coupling regime which enables reversible energy
exchange between photons and phonons. Meanwhile, quantum backaction heating
can pose a fundamental limit for backaction cooling. The solid curves ($A$,
$C$ and $E$) illustrate cooling processes associated with energy swapping,
counter-rotating-wave interaction and cavity dissipation, which one seeks to
enhance while suppressing heating for efficient mechanical motion cooling.

We focus on the resolved sideband regime $\kappa<\omega_{\mathrm{m}}$ and we
set $\Delta^{\prime}=-{\omega_{\mathrm{m}}}$, in which the beam splitter
interaction $a_{1}^{\dag}b_{1}+a_{1}b_{1}^{\dag}$ is on resonance. In this
case the dynamical stability condition from the Routh-Hurwitz criterion
\cite{StablePRA11} requires $2\left\vert G\right\vert <\omega_{\mathrm{m}}$.
To realize cooling, the cooperativity $C\equiv{4}\left\vert {G}\right\vert
^{2}/({\gamma\kappa})\gg1$ should also be satisfied. Starting from the master
equation, we obtain a set of differential equations for the mean values of the
second-order moments $\bar{N}_{a}=\langle{a_{1}^{\dag}a}_{1}\rangle$, $\bar
{N}_{b}=\langle{b_{1}^{\dag}b}_{1}\rangle$, $\langle{a{_{1}^{\dag}{{{b_{1}}}}%
}}\rangle$, $\langle{a{_{1}{{{b_{1}}}}}}\rangle$, $\langle{a{_{1}^{2}}}%
\rangle$ and $\langle{b{_{1}^{2}}}\rangle$ (see Supplementary Material
\cite{Supp}). In the steady state we obtain the phonon occupancy
\cite{SCPRL08,SCNJP08}
\begin{equation}
\bar{N}_{\mathrm{std}}\simeq\frac{{\gamma(4}\left\vert {G}\right\vert
^{2}+{\small \kappa}^{2}{)}}{{4}\left\vert {G}\right\vert ^{2}\left(
{\kappa+\gamma}\right)  }n_{\mathrm{th}}+\frac{{\kappa}^{2}+8\left\vert
{G}\right\vert ^{2}}{{16(\omega_{\mathrm{m}}^{2}-4}\left\vert {G}\right\vert
^{2})},\label{Nstd}%
\end{equation}
where the first term is the classical cooling limit and the second term
originates from the quantum backaction, consisting of both dissipation quantum
backaction related to the cavity dissipation (with the associated fluctuation-dissipation theorem)
and interaction quantum
backaction associated with the optomechanical interaction (see Supplementary
Material \cite{Supp} for full description). In the weak coupling regime, Eq.
(\ref{Nstd}) reduces to $\bar{N}_{\mathrm{std}}^{\mathrm{wk}}\simeq{\gamma
}n_{\mathrm{th}}/(\Gamma+\gamma)+{\kappa}^{2}/({16\omega_{\mathrm{m}}^{2})}$
with $\Gamma=4|{G}|^{2}/{\kappa}$, which agrees with Refs.
\cite{PRL07-1,PRL07-2}, and with $\kappa^{2}/(16\omega_{\mathrm{m}}^{2})$
the dissipation quantum backaction from fluctuation-dissipation. In the strong coupling regime, we obtain $\bar
{N}_{\mathrm{std}}^{\mathrm{str}}\simeq{\gamma}n_{\mathrm{th}}/({\kappa
+\gamma)}+|{G}|^{2}/[{2}({\omega_{\mathrm{m}}^{2}-4}|{G}|^{2})]$. In this case
the classical limit is restricted by the cavity dissipation rate $\kappa$,
while the interaction quantum backaction limit suffers from high coupling rate
$|{G}|$.

\begin{figure}[tb]
\centerline{\includegraphics[width=\columnwidth]{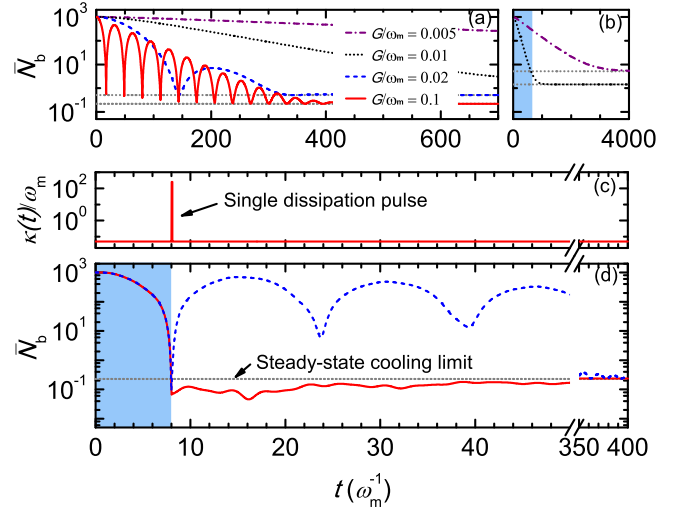}}
\caption{(color) (a) Time evolution of mean phonon number $\bar{N}_{b}$ for
$G/{\omega_{\mathrm{m}}}=0.005$, $0.01$, $0.02$ and $0.1$ (numerical results).
(b) $\bar{N}_{b}$ for $G/{\omega_{\mathrm{m}}}=0.005$ and $0.01$ with a wider
time interval. The shadowed region shows the same time interval with (a). (c)
Modulation scheme of the cavity dissipation rate ${\kappa(t)}$ for fast
cooling to the steady-state limit and (d) the time evolution of mean phonon
number $\bar{N}_{b}$ with (red solid curve) and without (blue dashed curve)
modulation for $G/{\omega_{\mathrm{m}}}=0.2$. Other parameters:
$n_{\mathrm{th}}=10^{3}$, $\gamma/{\omega}_{\mathrm{m}}=10^{-5}$,
${\kappa/\omega_{\mathrm{m}}=0.05}$. The dotted horizontal lines correspond to
the steady-state cooling limits, given by Eq. (\ref{Nstd}).}%
\label{Fig2}%
\end{figure}

To study the cooling dynamics beyond the steady state, we solve the
differential equations to obtain the time evolution of the mean phonon number
$\bar{N}_{b}$. For weak coupling, we have $\bar{N}_{b}^{\mathrm{wk}}\simeq
n_{\mathrm{th}}({\gamma+\Gamma e}^{-\Gamma t})/({\gamma+\Gamma})+[{\kappa}%
^{2}/({16\omega_{\mathrm{m}}^{2})](}1-{e}^{-\Gamma t}),$ which shows that the
mean phonon number decays exponentially with the cooling rate $\Gamma$. This
cooling rate is limited by the coupling strength, since in the cooling route
$A\rightarrow E$ the energy flow from the mechanical mode to the optical mode
(process $A$) is slower than the cavity dissipation (process $E$).

In the strong coupling regime, we obtain the time evolution of the mean phonon
number described by (see Supplementary Material \cite{Supp})
\begin{align}
\bar{N}_{b}^{\mathrm{str}}  &  =\bar{N}_{b,1}^{\mathrm{str}}+\bar{N}%
_{b,2}^{\mathrm{str}},\label{Nbstr}\\
\bar{N}_{b,1}^{\mathrm{str}}  &  \simeq n_{\mathrm{th}}\frac{{\gamma+\frac
{{1}}{2}e}^{-\frac{{\kappa+\gamma}}{2}t}\left[  {\kappa-\gamma}+({\kappa
+\gamma)}\cos({\omega}_{+}-{\omega}_{-})t\right]  }{{\kappa+\gamma}%
},\nonumber\\
\bar{N}_{b,2}^{\mathrm{str}}  &  \simeq\frac{\left\vert {G}\right\vert
^{2}\left[  1-{e}^{-\frac{{\kappa+\gamma}}{2}t}\cos({\omega}_{+}+{\omega}%
_{-})t\cos({\omega}_{+}-{\omega}_{-})t\right]  }{{2(\omega_{\mathrm{m}}^{2}%
-4}\left\vert {G}\right\vert ^{2})},\nonumber
\end{align}
where ${\omega}_{\pm}=\sqrt{{\omega_{\mathrm{m}}^{2}}\pm2|{G}|{\omega
_{\mathrm{m}}}}$ are the normal eigenmode frequencies. The phonon occupancy exhibits oscillation
under an exponentially-decaying envelope and can be divided
into two distinguished parts $\bar{N}_{b,1}^{\mathrm{str}}$ and $\bar{N}%
_{b,2}^{\mathrm{str}}$, where the first part originates from energy
exchange between optical and mechanical modes, and the second part is
induced by quantum backaction. $\bar{N}_{b,1}^{\mathrm{str}}$ reveals Rabi
oscillation with frequency $\sim2|{G}|$, whereas the envelopes have the
same exponential decay rate $\Gamma^{\prime}=({\kappa+\gamma})/2$ regardless
of the coupling strength $|{G}|$. This is because, in the strong coupling
regime, the cooling route $A\rightarrow E$ is subjected to the cavity
dissipation (process $E$), which has slower rate than the energy exchange
between phonons and photons (process $A$). This saturation
prevents a higher cooling speed for stronger coupling. In
Figs. \ref{Fig2}(a) and (b) we plot the numerical results based on the master
equation for various $G$. It shows that for weak coupling the cooling rate
increases rapidly as the coupling strength increases, whereas for strong
coupling the envelope decay no longer increases, instead the oscillation
frequency becomes larger.

\textit{Fast cooling to the steady-state limit.---}To speed up the cooling
process in the strong coupling regime, here we take advantage of high cavity
dissipation to\ dynamically strengthen the cooling process\ $E$. The internal
cavity dissipation is abruptly increased each time when the Rabi oscillation
reaches a minimum-phonon state, such as through RF-synchronized carrier
injection to the optical cavity \cite{xuNPhys2007}. At this time the system
has transited from state $|n,m\rangle$ to state $|n+1,m-1\rangle$. Once a
strong dissipation pulse is applied to the cavity so that the process $E$
dominates, the system will irreversibly transit from state $|n+1,m-1\rangle$
to state $|n,m-1\rangle$. The dissipation pulse has essentially behaves as a
switch to halt the reversible Rabi oscillation, resulting in the suppression
of the swap heating. To verify this dissipative cooling, in Figs. \ref{Fig2}
(c) and (d) we plot the modulation scheme and the corresponding time evolution
of mean phonon number $\bar{N}_{b}$ for ${\kappa/{\omega_{\mathrm{m}}}=0.05}$
and ${G/{\omega_{\mathrm{m}}}=0.2}$. At the end of the first half Rabi
oscillation cycle, $t\sim\pi/(2|{G}|)$, a dissipation pulse of pulsewidth
$0.01\pi/(2|{G}|)$ is applied. Detailed tradeoffs of the dissipation quantum
backaction and the interaction quantum backaction for varying pulsewidths are
shown in the Supplementary Material \cite{Supp}. After incidence of the
dissipation pulse, the phonon number reaches and remains near the steady-state
limit. For short time scales, the remaining small-amplitude oscillations mainly
originate from counter-rotating-wave interactions. Note that without
modulation (blue dashed curve), the steady-state cooling limit is reached only
after $t\simeq400/\omega_{\mathrm{m}}$; while with the modulation (red solid
curve), it only takes $t\simeq8/\omega_{\mathrm{m}}$ to cool below the same limit.

\textit{Breaking the fundamental limit of backaction cooling.} \textit{---}By
periodically modulating the cavity dissipation so as to continuously suppress
the swap heating, the phonon occupancy can be kept below the steady-state
cooling limit. Actually, each time after the dissipation pulse is applied, the
photon number quickly drops to the vacuum state, which equivalently
re-initializes the system. By periodic pulse application,
the system will periodically re-initializes, which keeps the
phonon occupancy in an instantaneous-state cooling limit as verified in
Fig. \ref{Fig3}. The
instantaneous-state cooling limit is given by (see Supplementary Material
\cite{Supp})
\begin{equation}
\bar{N}_{\mathrm{ins}}\simeq\frac{\pi{\gamma}n_{\mathrm{th}}}{4\left\vert
{G}\right\vert }+\frac{\pi^{2}\left\vert {G}\right\vert ^{4}}{{(\omega
_{\mathrm{m}}^{2}-}\left\vert {G}\right\vert ^{2}){(\omega_{\mathrm{m}}^{2}%
-4}\left\vert {G}\right\vert ^{2})}. \label{Nins}%
\end{equation}
Here the first term comes from $\bar{N}_{b,1}^{\mathrm{str}}$ for $t\simeq
\pi/(2|{G}|)$, which shows a $\pi{\kappa/(}4|{G}|)$ times reduction of
classical steady-state cooling limit. The second term of $\sim\pi
^{2}\left\vert {G}\right\vert ^{4}/{\omega_{\mathrm{m}}^{4}}$, obtained from
$\bar{N}_{b,2}^{\mathrm{str}}$ when $t\simeq\pi/{\omega_{\mathrm{m}}}$,
reveals that the second order term of $|{G}|/{\omega_{\mathrm{m}}}$ in quantum
backaction has been removed in our approach, leaving only the higher-order
terms. Note that the cooling limit (\ref{Nins}) is the sum of the
individual minimum of $\bar{N}_{b,1}^{\mathrm{str}}$ and $\bar{N}%
_{b,2}^{\mathrm{str}}$ in their first oscillation cycle. Notably, in Fig.
\ref{Fig3} we demonstrate that the modulation is switchable. If we turn on the
modulation (\textquotedblleft ON\textquotedblright\ region), the system will
reach the instantaneous-state cooling limit; if we turn off the modulation
(\textquotedblleft OFF\textquotedblright\ region), the system transits back to
the steady-state cooling limit.

\begin{figure}[tb]
\centerline{\includegraphics[width=\columnwidth]{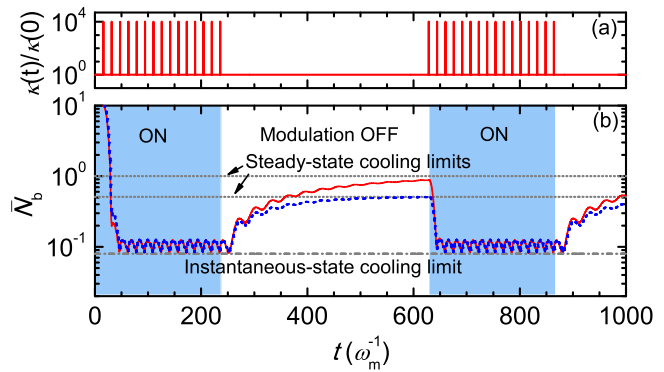}}
\caption{(color) Modulation scheme of ${\kappa(t)/\kappa(0)}$ (a) and the
corresponding time evolution of $\bar{N}_{b}$ (b) for $G/{\omega_{\mathrm{m}}%
}=0.1$, ${\kappa(0)/\omega_{\mathrm{m}}=0.01}$ (red solid curve) and $0.02$
(blue dashed curve). In (b), the two dotted horizontal lines (from top to
bottom) denoting the respective steady-state cooling limits depending on the
cavity decay ${\kappa(0)}$, given by Eq. (\ref{Nstd}); the dash-dotted line
denotes the instantaneous-state cooling limit independent of ${\kappa(0)}$,
given by Eq. (\ref{Nins}); the \textquotedblleft ON\textquotedblright\ and
\textquotedblleft OFF\textquotedblright\ regions corresponds that the
modulation is turned on and off, respectively; the vertical coordinate range
from $10$ to $10^{3}$ is not shown. Other parameters: $n_{\mathrm{th}}=10^{3}%
$, $\gamma/{\omega}_{\mathrm{m}}=10^{-5}$.}%
\label{Fig3}%
\end{figure}

In particular, from Eq. (\ref{Nbstr}), the interaction quantum backaction
heating term $\bar{N}_{b,2}^{\mathrm{str}}$ forms a carrier-envelope type
evolution, where the carrier oscillation represents the counter-rotating-wave
interaction and the envelope oscillation is a result of coherent energy
exchange due to strong coupling. The minimum of $\bar{N}_{b,2}^{\mathrm{str}}$
is dependent on the carrier-envelope frequency matching. If $({\omega}%
_{+}+{\omega}_{-})/({\omega}_{+}-{\omega}_{-})=k$ ($k=3,5$...), yielding
$|{G}|/{\omega_{\mathrm{m}}=k/(k}^{2}+1)=0.3$, ${5/26}$..., $\bar{N}%
_{b,2}^{\mathrm{str}}$ reaches a minimum $\sim\frac{\pi{\kappa}|{G}%
|}{{8(\omega_{\mathrm{m}}^{2}-4}|{G}|^{2})}$ for $t\simeq\pi/(2|{G}|)$. Here we obtain the optimized instantaneous-state cooling limit
as (\cite{Supp})
\begin{equation}
\bar{N}_{\mathrm{ins}}^{\mathrm{opt}}\simeq{\frac{\pi{\kappa}}{4\left\vert
{G}\right\vert }}\left[  \frac{{\gamma}n_{\mathrm{th}}}{{\kappa}}%
+\frac{\left\vert {G}\right\vert ^{2}}{{2(\omega_{\mathrm{m}}^{2}-4}\left\vert
{G}\right\vert ^{2})}\right]  , \label{Ninsmat}%
\end{equation}
which reduces both the classical and quantum steady-state cooling limits by a
factor of $\pi{\kappa/(}4|{G}|)$. Remarkably, this reduction is significant
when the system is in the deep strong coupling regime. Besides, the leading
order of the interaction quantum backaction heating scales as ${\kappa}%
|{G}|/{\omega_{\mathrm{m}}^{2}}$, which can be a few orders of magnitude lower
than the steady-state case, representing large suppression of quantum backaction.

\begin{figure}[tb]
\centerline{\includegraphics[width=\columnwidth]{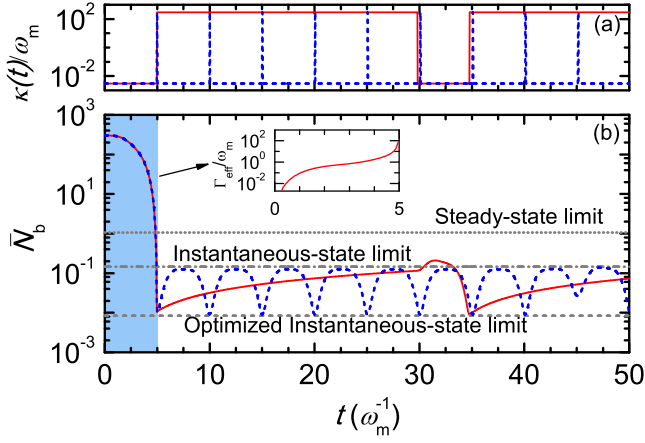}}
\caption{(color) Different pulsewidth (a) and the corresponding time evolution
(b) of $\bar{N}_{b}$. The horizontal lines indicate the three cooling limits
given by Eqs. (\ref{Nstd}), (\ref{Nins}) and (\ref{Ninsmat}) from top to
bottom. The inset shows the effective cooling rate as a function of time.
Parameters : $n_{\mathrm{th}}=300$, ${G}/{\omega_{\mathrm{m}}=0.3}$,
${\kappa/\omega_{\mathrm{m}}=0.003}$ and $\gamma/{\omega}_{\mathrm{m}}%
=10^{-5}$.}%
\label{Fig4}%
\end{figure}

To verify suppression of the interaction quantum backaction heating, in Fig.
\ref{Fig4} we plot the cooling dynamics with dissipation modulation for
$G/{\omega_{\mathrm{m}}}=0.3$ and ${\kappa/{\omega_{\mathrm{m}}}=0.003}$. The single modulation pulse brings down the phonon occupation to the
optimized instantaneous-state cooling limit described in Eq. (\ref{Ninsmat}),
with the time-dependent effective cooling rate ${\Gamma}_{\mathrm{eff}}%
=(d\bar{N}_{b}/dt)/\bar{N}_{b}$ shown in the inset. With short-pulse
modulation (blue dashed curve), the remaining oscillation, mainly induced by
the counter-rotating-wave interaction, has a quasi-period of $\pi/(2|{G}|)$
due to frequency matching. This small-amplitude fluctuations around the
instantaneous-state cooling limit might affect future quantum protocols, but
in the sense of time-averaged through timescales larger than $\pi/(2|{G}|)$,
the cooling limit can be viewed stable. By using long-pulse modulation, the
quasi-periodic fluctuations can be suppressed (red solid curve). This is
because the large dissipation suppresses the interaction quantum backaction.
The cost is that the dissipation quantum backaction takes effect and gradually
increases the phonon number. This trade-off can be balanced by optimizing the
pulsewidth as shown in the Supplementary Material \cite{Supp}.

Figure \ref{Fig5} plots the cooling limits as functions of ${G}/{\omega
_{\mathrm{m}}}$, which reveals that instantaneous-state cooling limits are
much lower than steady-state cooling limits. For small coupling rates, we
observed that interaction quantum backaction is insignificant and suppression
of swap heating is the main origin of cooling limit reduction. For large
coupling rates, suppressing interaction quantum backaction is crucial for
obtaining lower limits. Typically, the cooling limits can be reduced by a few
orders of magnitude. For example, when $G/{\omega_{\mathrm{m}}}=0.3$ and
${\kappa/{\omega_{\mathrm{m}}}=0.003}$, we obtain $\bar{N}_{\mathrm{std}}%
=3.4$, while $\bar{N}_{\mathrm{ins}}^{\mathrm{opt}}=0.03$, corresponding to
more than $100$ times of phonon number suppression.

\begin{figure}[tb]
\centerline{\includegraphics[width=7.5cm]{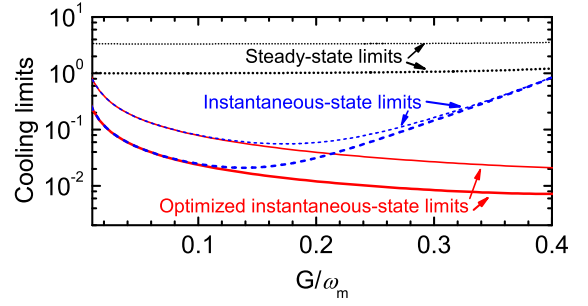}}
\caption{(color) Cooling limits given by Eqs. (\ref{Nstd}) (black dotted
curves), (\ref{Nins}) (blue dashed curves) and (\ref{Ninsmat}) (red solid
curves) versus ${G}/{\omega_{\mathrm{m}}}$ for $n_{\mathrm{th}}=10^{3}$ (thin
curves), $3\times10^{2}$ (thick curves). Other parameters are ${\kappa
/\omega_{\mathrm{m}}=0.003}$ and $\gamma/{\omega}_{\mathrm{m}}=10^{-5}$.}%
\label{Fig5}%
\end{figure}

Experimentally, the dynamic control of cavity dissipation can be realized, for
example, by modulating free-carrier plasma density
\cite{xuNPhys2007,PRL2013Baba,AddIEEE87} or using a light absorber/scatterer
\cite{AddNJP08}. Note that we assume $G$ is kept unchanged when the
dissipation pulses are applied, which corresponds to the invariableness of the
intracavity field ${\alpha}$. This can be fulfilled by simultaneously changing
the driving ${\Omega(t)}$, so that equation $[{i\Delta}^{\prime}%
-\kappa(t){/2]\alpha-{i\Omega(t)=0}}$ is satisfied all the time (\cite{Supp}; Section VI). Here modulated square-shaped dissipation
pulses are used though further simulations show that the results are
irrespective of the pulse shape, as long as they are executed quickly with
strong enough peak value at the desired time. This is because the pulse dissipation mainly relies on the pulse area.

In summary, we examined cooling of mesoscopic mechanical resonators
in the strong coupling regime and propose dynamic dissipative schemes which
possess large cooling rates, low cooling limits, and long-time stability. By making use of the cavity dissipation, swap heating
can be strongly avoided and the interaction\ quantum backaction largely
suppressed, with great advantages over the current conventional
cooling approaches. For example, a single dissipation pulse enables more than
$50$ times higher cooling rate; with periodic modulation of cavity
dissipation, the cooling limit can be reduced by more than two orders of
magnitude. Different from the cooling schemes with modulated coupling
\cite{PulPRB09,PulPRA11-1,PulPRA11-2,PulPRL11,PulPRL12}, we take advantage of
large cavity dissipation, usually regarded as a noise source.
Together with recent proposals of other dissipative effects such as
two-level ensembles \cite{AddPRA09} or photothermal effect
\cite{AddCRP11,AddPRA11}, we demonstrate that cavity
dissipation (even in the presence of the considered dissipation quantum
backaction) can be viewed as a resource. Compared with the dissipative
coupling \cite{DCPRL09,DCPRL11,LiPRL2009}, this active dissipation control does not require the coupling between the cavity
dissipation and the mechanical resonator. The dynamic dissipative cooling
provides a new way for exploring the quantum regime of mechanical devices,
ranging from mechanical ground state preparation, to generation of
mesoscopic quantum states, and quantum-limited measurements.

\begin{acknowledgments}
We thank H.-K. Li for discussions. This work is supported by DARPA ORCHID program (C11L10831), 973
program (2013CB328704 and 2013CB921904), NSFC (11004003, 11222440, and 11121091), and
RFDPH (20120001110068). Y.C.L is supported by the PhD Students'
Short-term Overseas Research Program of Peking University and Scholarship
Award for Excellent Doctoral Candidates.
\end{acknowledgments}

\end{document}